\documentclass[aps, prd, twocolumn, floatfix, nofootinbib, superscriptaddress]{revtex4-1}

\usepackage{amsmath,amsfonts,amssymb,bm}
\usepackage{dsfont}
\usepackage{graphicx}
\usepackage{color}
\usepackage{soul}

\usepackage[mathscr,scaled=1.15]{urwchancal}
\DeclareFontFamily{OT1}{pzc}{}
\DeclareFontShape{OT1}{pzc}{m}{it}%
{<-> s * [1.15] pzcmi7t}{}
\DeclareMathAlphabet{\mathpzc}{OT1}{pzc}{m}{it}

\definecolor{purple}{rgb}{0.5,0,0.5}
\definecolor{blue}{rgb}{0.0,0,0.9}
\definecolor{prdblue}{rgb}{0.133,0.118,0.498}
\usepackage[colorlinks=true, pdfstartview=FitV, linkcolor=prdblue, citecolor= prdblue, urlcolor=prdblue]{hyperref}

\begin{document}

\title{Pion and kaon valence-quark quasiparton distributions}

\author{Shu-Sheng Xu}
\affiliation{Department of Physics, Nanjing University, Nanjing, Jiangsu 210093, China}

\author{Lei Chang}
\email[]{leichang@nankai.edu.cn}
\affiliation{School of Physics, Nankai University, Tianjin 300071, China}

\author{Craig D. Roberts}
\email[]{cdroberts@anl.gov}
\affiliation{Physics Division, Argonne National Laboratory, Argonne, Illinois
60439, USA}

\author{Hong-Shi Zong}
\affiliation{Department of Physics, Nanjing University, Nanjing, Jiangsu 210093, China}
\affiliation{Joint Center for Particle, Nuclear Physics and Cosmology, Nanjing, Jiangsu 210093, China}

\date{22 February 2018}

\begin{abstract}
Algebraic \emph{Ans\"atze} for the Poincar\'e-covariant Bethe-Salpeter wave functions of the pion and kaon are used to calculate their light-front wave functions (LFWFs), parton distribution amplitudes (PDAs), quasi-PDAs (qPDAs), valence parton distribution functions (PDFs), and quasi-PDFs (qPDFs).
The LFWFs are broad, concave functions; and the scale of flavour-symmetry violation in the kaon is roughly 15\%, being set by the ratio of emergent masses in the $s$-and $u$-quark sectors.
qPDAs computed with longitudinal momentum $P_z =1.75\,$GeV provide a semiquantitatively accurate representation of the objective PDA; but even with $P_z=3\,$GeV, they cannot provide information about this amplitude's endpoint behaviour.
On the valence-quark domain, similar outcomes characterise qPDFs.  In this connection, however, the ratio of kaon-to-pion $u$-quark qPDFs is found to provide a good approximation to the true PDF ratio on $0.3\lesssim x \lesssim 0.8$, suggesting that with existing resources computations of ratios of quasi-parton-distributions can yield results that support empirical comparison.
\end{abstract}



\maketitle

\section{Introduction}
Since the discovery of quarks in deep inelastic scattering (DIS) experiments at the Stanford Linear Accelerator Center \cite{Taylor:1991ew, Kendall:1991np, Friedman:1991nq}, parton distributions have occupied a central role in high-energy nuclear and particle physics; and today there is a vast international programme aimed at their measurement.  Such measurements are possible owing to the existence of factorisation theorems \cite{Collins:1989gx}, which entail that the cross-sections for various processes can be written as the convolution of a piece calculable using QCD perturbation theory and a parton distribution function (PDF), which is independent of the process used.  PDFs are therefore a characterising property of the chosen hadronic target.  This means they are also essentially nonperturbative, \emph{i.e}.\ their calculation is a problem in strong QCD (sQCD), with sound results demanding the use of a nonperturbative method with a traceable connection to QCD.  The extent and importance of this computational challenge is canvassed elsewhere \cite{Holt:2010vj, Brodsky:2015aia, Horn:2016rip}.

An \emph{ab initio} approach to sQCD is provided by the numerical simulation of lattice-regularised QCD (lQCD).   However, a given PDF is mathematically defined as an expectation value of some bilocal operator evaluated along a light-like line, an object which cannot be evaluated using the methods of lQCD.  This approach only provides access to the expectation value of local operators, \emph{i.e}., in this context, to the Mellin moments of the PDF.  That would not be an issue if every moment were accessible because a probability distribution is completely determined once all its moments are known.  However, discretised spacetime does not possess the full rotational symmetries of the Euclidean continuum.  Hence, only the lowest three non-trivial moments can readily be calculated; and they are insufficient to support a model-independent reconstruction of the PDF.  A number of paths are being pursued to circumvent this problem \cite{Liu:1993cv, Ji:2013dva, Radyushkin:2016hsy, Radyushkin:2017gjd, Chambers:2017dov}.

Herein, using continuum methods in quantum field theory, we explore some aspects of the large $P_z$ (longitudinal-momentum) approach to the lQCD computation of parton distributions \cite{Ji:2013dva}, \emph{viz}.\ since the maximum value of $P_z$ is bounded in any lQCD simulation, what is the lowest value of $P_z^{\rm max}$ for which the quasi-distribution provides a realistic sketch of the true distribution; and given a $P_z^{\rm max}$ quasi-distribution, is it possible to extract reliable information about the true distribution?
These issues are considered, \emph{e.g}.\ in Refs.\,\cite{Gamberg:2015opc, Bacchetta:2016zjm} using spectator models of the proton.
We, on the other hand, choose to focus on the pion and kaon because there has been significant progress in the continuum computation of the distribution amplitudes and functions of these systems in recent years \cite{Brodsky:2006uqa, Chang:2013pq, Cloet:2013tta, Chang:2013nia, Segovia:2013eca, Shi:2014uwa, Chang:2014lva, Mezrag:2014jka, Shi:2015esa, Raya:2015gva, Chen:2016sno, Mezrag:2016hnp, Li:2016dzv, Raya:2016yuj, Gao:2016jka, Qin:2017lcd, Gao:2017mmp, deTeramond:2018ecg} and first lQCD results on some of their quasi-distributions are now available \cite{Zhang:2017bzy, Chen:2017gck}.   Our discussion complements the analysis in Refs.\,\cite{Nam:2017gzm, Broniowski:2017wbr}.

Sec.\,\ref{SecModel} describes the framework used to represent $\pi$- and $K$-mesons as bound states in quantum field theory, specifying the elements at an hadronic scale, $\zeta_H \sim 1\,$GeV.  Herein, we do not consider perturbative QCD evolution \cite{Dokshitzer:1977, Gribov:1972, Lipatov:1974qm, Altarelli:1977, Lepage:1979zb, Efremov:1979qk, Lepage:1980fj}: such evolution doesn't affect our comparisons between quasi-distributions and their associated light-front distribution functions, which are all made at the same scale.  Section~\ref{SecLFWF} focuses on $\pi$ and $K$ leading-twist light-front wave functions, their derived parton distribution amplitudes and attendant quasi-distributions, introducing general formulae and providing numerical illustrations.  The kindred analysis of valence-dressed-quark quasi-parton-distribution-functions (qPDFs) is presented in Sec.\,\ref{SecqPDFSs}.  We conclude in Sec.\,\ref{secEpilogue}.


\section{Pion and Kaon Bound States}
\label{SecModel}
Many insights into the character of the pion and kaon have been drawn using the following simple expressions for the relevant dressed-quark propagators and Bethe-Salpeter amplitudes \cite{Chang:2013pq, Chang:2014lva, Mezrag:2014jka, Chen:2016sno}:
\begin{subequations}
\label{Algebraic}
\begin{align}
S_f(k) & =[ -i \gamma\cdot k + M_f] \Delta(k^2,M_f^2)\,,\\
{\mathpzc n}_G \Gamma_G(k;P_G) & = i \gamma_5  \int_{-1}^1\,dw\,\rho_G(w) \hat \Delta(k_w^2,\Lambda^2)\,,
\label{EqGamp}
\end{align}
\end{subequations}
where
$M_f$ is the dressed-quark mass evaluated in the neighbourhood $k^2\simeq 0$,  $f=u,s$ (we work in the isospin symmetric limit, so $M_u=M_d$);
$\Delta(s,t) = 1/[s+t]$, $\hat \Delta(s,t) = t \Delta(s,t) $;
$k_w = k+ (w/2) P$, with $P^2=-m_G^2$, $G=\pi, K$;
and $\rho_{G=\pi,K}(w)$ is a spectral weight whose form determines the pointwise behaviour of the associated meson's Bethe-Salpeter amplitude, with ${\mathpzc n}_G$ the related canonical normalisation constant.
One of the strengths of these \emph{Ans\"atze} is that they can be chosen to ensure that a primarily algebraic computation yields results which are pointwise similar to the most sophisticated predictions currently available for parton distribution amplitudes and functions, PDAs and PDFs.
Notably, our approach to the continuum bound-state problem is Poincar\'e-covariant and hence, with complete generality, we may write
\begin{equation}
P_G = (0,0,P_z, i E_P)\,,\quad E_P = [P_z^2 + m_G^2]^{\tfrac{1}{2}}.
\end{equation}

One branch of our analysis will focus on the leading-twist two-dressed-parton distribution amplitudes of the $\pi$- and $K$-mesons, the computation of which requires a projection onto the light-front of the given meson's (unamputated) Bethe-Salpeter wave function.  Working with the $K^+$ meson as an illustration, this wave function is ($k^K_- = k- P_K/2$)
\begin{align}
\chi_K(k^K_-;P_K) & = S_u(k) \Gamma_K(k^K_-;P_K) S_s(k-P)\,;
\end{align}
and the part which contributes to the leading-twist (twist-two) PDA is readily computed:
\begin{subequations}
\begin{align}
{\mathpzc n}_K \chi_K^{(2)}(k^K_-;P_K) & =
{\mathpzc M}(k;P_K)
\int_{-1}^1\,dw\,\rho_K(w) {\mathpzc D}(k;P_K)\,,\\
\nonumber
{\mathpzc M}(k;P_K)  & = -\gamma_5[ \gamma\cdot P_K  M_{u}+ \gamma\cdot k (M_u-M_s) \\
& \qquad + \sigma_{\mu\nu} k_\mu P_{K\nu}] \,,\\
\nonumber {\mathpzc D}(k;P_K) & =  \Delta(k^2,M_u^2) \Delta((k-P)^2,M_s^2) \\
 & \qquad \times \hat\Delta(k_{w-1}^2,\Lambda^2)\,.
\end{align}
\end{subequations}

One may now introduce two Feynman parameters, combine the denominators into a single quadratic form, and thereby arrive at:
\begin{align}
\label{X2a}
\chi_K^{(2)}(k^K_-;P_K) & = {\mathpzc M}(k;P_K) \int_0^1 \,d\alpha\,2 \, {\mathpzc X}_K(\alpha;\sigma^3(\alpha))\,,
\end{align}
with $\sigma = (k-\alpha P_K)^2+ \Omega_K^2$,
\begin{align}
\nonumber
\Omega_K^2 & = v M_u^2 + (1-v)\Lambda^2 \\
\nonumber
& + (M_s^2-M_u^2)\left(\alpha - \tfrac{1}{2}[1-w][1-v]\right) \\
&  + ( \alpha [\alpha-1] + \tfrac{1}{4} [1-v] [1-w^2]) M_K^2\,,
\label{Omega}\\
\nonumber
{\mathpzc X}_K(\alpha;\sigma^3) & =
\left[
\int_{-1}^{1-2\alpha} \! dw \int_{1+\frac{2\alpha}{w-1}}^1 \!dv \right. \\
&\quad  \left. + \int_{1-2\alpha}^1 \! dw \int_{\frac{w-1+2\alpha}{w+1}}^1 \!dv \right]\frac{\rho_K(w) }{{\mathpzc n}_K } \frac{\Lambda^2}{\sigma^3}\,.\label{X2c}
\end{align}
Formulae for the $\pi$-meson are readily obtained by setting $s\to d$, $m_K \to m_\pi$, and using isospin symmetry.

As has long been known \cite{Pagels:1974se} and is demonstrated for parton distributions in, \emph{e.g}.\ Refs.\,\cite{Segovia:2013eca, Shi:2014uwa, Shi:2015esa, Li:2016dzv, Li:2016mah}, distinctions between the $K$- and $\pi$-mesons are driven by dynamical chiral symmetry breaking (DCSB), expressed in Eq.\,\eqref{Omega} by the difference between the dressed $s$- and $u$ quark masses: $(M_s^2-M_u^2)$.


%
\section{Light-Front Wave Functions and Quasi-PDAs}
\label{SecLFWF}
\subsection{Algebraic Analysis}
\label{SecqPDAs}
The pseudoscalar meson's leading-twist two-dressed-parton light-front wave function (LFWF) can be written:
\begin{align}
\label{EqLFWFG}
f_K \psi_K^{\uparrow\downarrow}(x,k_\perp^2) & = {\rm tr}_{\rm CD}
\int_{dk_\parallel} \delta_n^x(k_K)
\gamma_5 \gamma\cdot n \chi_K^{(2)}(k^K_-;P_K)\,,
\end{align}
where
$f_K$ is the kaon's leptonic decay constant;
the trace is over colour and spinor indices;
$\int_{dk_\parallel} = (1/\pi)\int dk_3 dk_4$;
$\delta_n^x(k_K)= \delta(n\cdot k - x n\cdot P_K)$;
and $n$ is a light-like four-vector, $n^2=0$, $n\cdot P_K = -m_K$.  The twist-two PDA follows immediately:
\begin{subequations}
\begin{align}
\label{EqVarphi}
\varphi_K(x) & = \frac{1}{16\pi^3}\int d^2k_\perp  \psi_K^{\uparrow\downarrow}(x,k_\perp^2)\,,\\
& \int_0^1dx\,\varphi_K(x)  = 1\,.
\end{align}
\end{subequations}

Consider now the following Mellin moments:
\begin{subequations}
\begin{align}
\label{EqL1}
 \langle x^m &\rangle_{\Psi^{\uparrow\downarrow}_K} =   \int_0^1 dx \, x^m\,  \psi_K^{\uparrow\downarrow}(x,k_\perp^2)\\
& = \frac{1}{f_K n\cdot P}\int_{dk_\parallel} \left[\frac{n\cdot k}{n\cdot P}\right]^m \gamma_5 \gamma\cdot n \chi_K^{(2)}(k^K_-;P_K)\\
& = \frac{12}{f_K} \int_0^1 d\alpha \, \alpha^m\, {\mathpzc Y}_K(\alpha;\sigma^2)\,,\label{EqL3}
\end{align}
\end{subequations}
where we have used Eqs.\,\eqref{X2a}--\eqref{X2c}, and ${\mathpzc Y}(\alpha;\sigma_\perp^2) = [M_u(1-\alpha) +M_s\alpha ]{\mathpzc X}(\alpha;\sigma_\perp^2)$, $\sigma_\perp=k_\perp^2+\Omega_K^2$.  Comparing Eqs.\,\eqref{EqL1} and \eqref{EqL3}, it is apparent that
\begin{align}
\label{EqLFWF}
\psi_K^{\uparrow\downarrow}(x,k_\perp^2) = \frac{12}{f_K}  {\mathpzc Y}_K(x;\sigma_\perp^2)\,,
\end{align}
where $\alpha \to x$ in Eqs.\,\eqref{Omega}, \eqref{X2c}.  The $\pi$-meson formula is obvious by analogy.  The compactness of these results is one merit of the algebraic \emph{Ans\"atze} in Eqs.\,\eqref{Algebraic}.

Combining Eqs.\,\eqref{EqLFWFG}, \eqref{EqVarphi}, a twist-two dressed-parton quasi-PDA (qPDA) is obtained via the replacement $n \to \tilde n$, $\tilde n=(0,0,1,0)$, \emph{viz}.
\begin{align}
f_K\, \tilde\varphi_K(x) = {\rm tr}_{\rm CD} \int_{dk} \!\!
\delta_{\tilde n}^x(k_K) \,\gamma_5\gamma\cdot \tilde n\,  \chi_K^{(2)}(k^K_-;P_K)\,,
\end{align}
where $\delta_{\tilde n}^x(k_K)= \delta(\tilde n\cdot k - x \tilde n\cdot P_K)$; and $\int_{dk}$ is a Poincar\'e-invariant definition of the four-dimensional integral.  Following a series of steps similar to those used above, one arrives directly at the following result:
\begin{align}
\label{EqqPDA}
 \tilde\varphi_K(x) & = \frac{P_z}{16\pi^3} \int_0^1 d\alpha \int_{-\infty}^\infty \! dk \,\psi_K^{\uparrow\downarrow}(\alpha, k^2+(x-\alpha)^2 P_z^2)\,.
\end{align}
The expression for $\tilde\varphi_\pi(x)$ is obvious by analogy and matches Eq.\,(20) in Ref.\,\cite{Radyushkin:2017gjd}.

\subsection{Numerical Illustrations}
\subsubsection{Wave Functions}
%
It is now possible to study the $P_z$-evolution of the pointwise-form of meson qPDAs and chart their connection with the objective PDA.  To proceed, it is necessary to specify the parameters and spectral densities in Eq.\,\eqref{Algebraic}.  For the latter, we use
\begin{align}
\nonumber
{\mathpzc u}_G\, \rho_G(w)  & = \frac{1}{2 b_0^G} \left[
{\rm sech}^2([w-w_0^G]/[2 b_0^G]) \right. \\
& \quad  \left . + {\rm sech}^2([w+w_0^G]/[2 b_0^G]) \right] [1 +w {\mathpzc v}_G ] \,,
\label{Eqrho}
\end{align}
where $b_0^G$, $w_0^G$, ${\mathpzc v}_G$, are parameters, and ${\mathpzc u}_G$ is a derived constant that ensures unit normalisation of the density.  This form is compact and yet has sufficient flexibility to produce pion and kaon valence-quark PDAs and PDFs whose features are consistent with contemporary predictions.

Regarding the parameters, we choose $M_u=0.31\,$GeV, matching the infrared scale of the $u$-quark mass function obtained using modern gap-equation kernels \cite{Binosi:2016wcx}; set $M_s= 1.2 M_u$, which is typical of the size obtained in phenomenologically efficacious continuum analyses \cite{Maris:1997tm, Shi:2015esa}; float $\Lambda_{\pi,K}$ to fit the leptonic decay constants:
\begin{subequations}
\label{fHformulae}
\begin{align}
&f_\pi = \frac{1}{n\cdot P_\pi} {\rm tr}_{\rm CD} \! \int_{dk}\!\!\!
\gamma_5 \gamma\cdot n \chi_\pi (k_-^\pi;P_\pi)\,,\\
& f_K = \frac{1}{n\cdot P_K} {\rm tr}_{\rm CD}\int_{dk}
\gamma_5 \gamma\cdot n \chi_K (k_-^K;P_K)\,,
\end{align}
\end{subequations}
where $m_\pi=0.14\,$GeV, $m_K=0.49\,$GeV; and choose $b_0^G$, $w_0^G$  such that the meson PDAs are broad, concave functions whose lowest nontrivial Mellin moments match those obtained in modern analyses \cite{Horn:2016rip}:
\begin{subequations}
\label{EqMoments}
\begin{align}
\langle (2x-1)^2\rangle_{\varphi_\pi} & := \int_0^1 dx\, (2x-1)^2 \varphi_\pi(x) \approx 0.25\,,\\
\langle 2x-1\rangle_{\varphi_K} & \approx -0.04\,,\; \langle (2x-1)^2\rangle_{\varphi_K} \approx 0.25\,.
\end{align}
\end{subequations}
With
\begin{equation}
\label{parameters}
\begin{array}{cccc|cccc}
\Lambda_\pi  & b_0^\pi & w_0^\pi & {\mathpzc v}_\pi & \Lambda_K & b_0^K & w_0^K & {\mathpzc v}_K \\ \hline
M_u & 0.1 & 0.73 & 0 & 2 \Lambda_\pi  & b_0^\pi & 0.95 & 0.16
\end{array}
\end{equation}
we obtain $f_\pi =0.092\,$GeV, $f_K = 0.11\,$GeV, in agreement with experiment \cite{Olive:2016xmw}, and satisfy Eqs.\,\eqref{EqMoments}.  Recall that in connection with quantities that undergo QCD evolution, our models should be understood as producing results valid at an hadronic scale, $\zeta_H \sim 1\,$GeV.

\begin{figure}[t]
\begin{center}
\includegraphics[clip, width=0.4\textwidth]{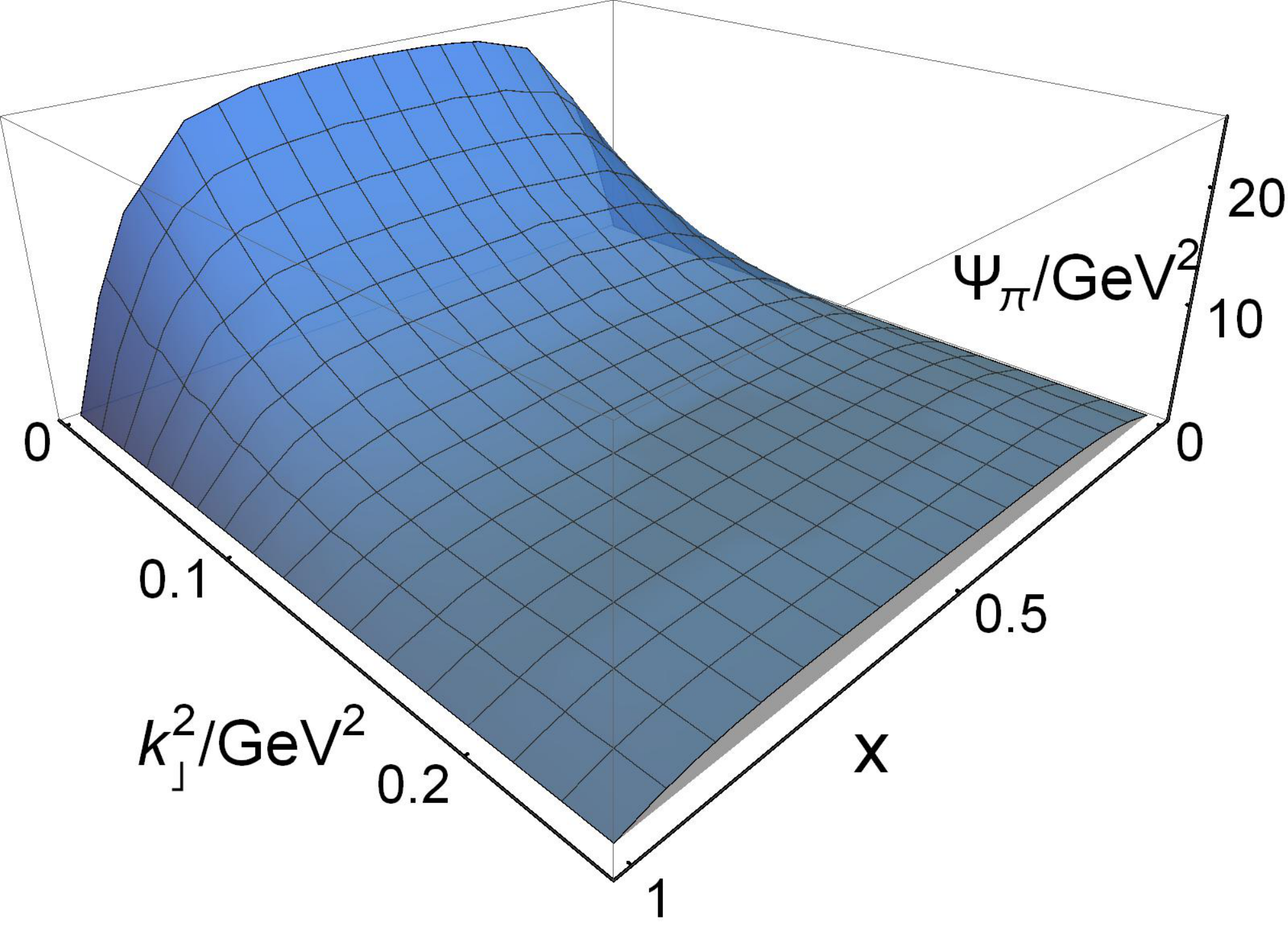}
\includegraphics[clip, width=0.4\textwidth]{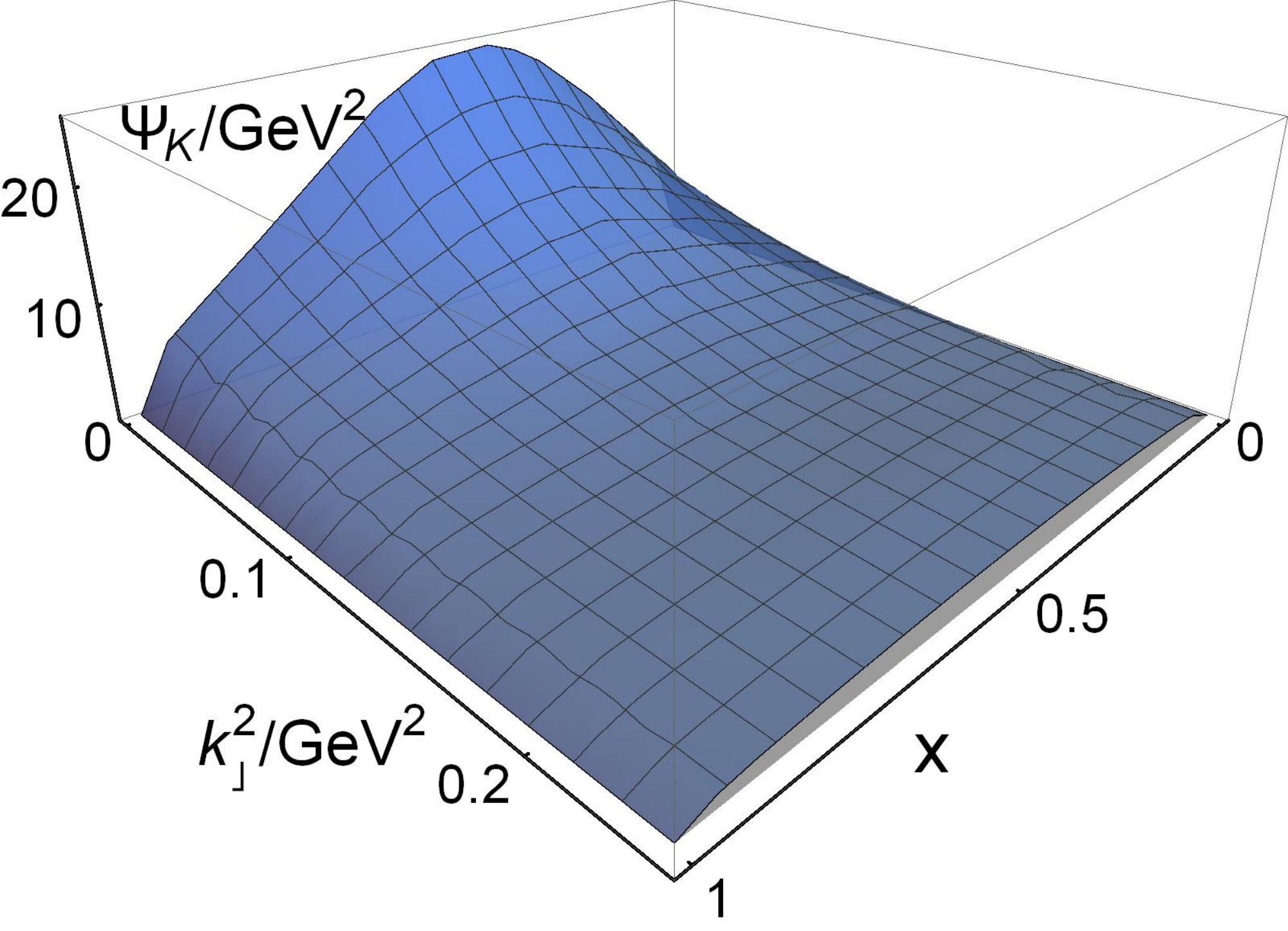}
\end{center}
\caption{\label{FigLFWF}
Leading-twist two-dressed-parton light-front wave functions of the pion (upper panel) and kaon (lower panel).  Each is normalised such that $\int dx d^2k_\perp \psi^{\uparrow\downarrow}(x,k_\perp^2) = 1 $.
}
\end{figure}

The pion and kaon leading-twist dressed-parton LFWFs, obtained using Eqs.\,\eqref{Algebraic}, \eqref{EqLFWF}, \eqref{Eqrho}, \eqref{parameters}, are depicted in Fig.\,\ref{FigLFWF}.
Considered as a function of $x$, with $k_\perp^2$ fixed, these wave functions are broad and concave.  Conversely, at fixed $x$, they fall as $1/k_\perp^4$ on $k_\perp^2\gg \Lambda_G^2$.
%
In QCD, the behaviour is $1/k_\perp^2$ (up to $\ln k_\perp^2$-corrections).  Our model's decay rate is amplified because Eq.\,\eqref{EqGamp} retains only the $\gamma_5$ piece of the pseudoscalar meson Bethe-Salpeter amplitude.  Two additional ``pseudovector'' Dirac structures are prominent in symmetry-preserving solutions of the Bethe-Salpeter equations for light pseudoscalars \cite{Maris:1997tm}; and the omission of these components produces the $1/k_\perp^4$ decay at ultraviolet momenta \cite{Maris:1998hc}.  This has a benefit: all integrals appearing herein are convergent.  Restoring the pseudovector components, the LFWFs recover the $1/k_\perp^2$ decay characteristic of meson wave functions in QCD.  Consideration of regularisation and renormalisation is then necessary; but that is straightforward and has no material effect on our discussion, which is why we exploit the simplicity of Eq.\,\eqref{EqGamp}.

The lower panel of Fig.\,\ref{FigLFWF} reveals that the $K^+$ LFWF is distorted, with its maximum located at $(x=0.44,k_\perp^2=0)$, \emph{viz}.\ displaced relative to that of $\psi_\pi$ and thereby indicating that the dressed $\bar s$-quark carries a larger fraction of the kaon's momentum than the $u$-quark.  As noted elsewhere \cite{Segovia:2013eca, Shi:2014uwa, Shi:2015esa, Li:2016dzv, Li:2016mah}, the magnitude of this $SU(3)$-flavour-symmetry breaking shift ($\simeq 15$\%) is set by DCSB mass-scales, as expressed, \emph{e.g}. in $M_s/M_u = 1.2$.

\begin{figure}[t]
\begin{center}
\includegraphics[clip, width=0.42\textwidth]{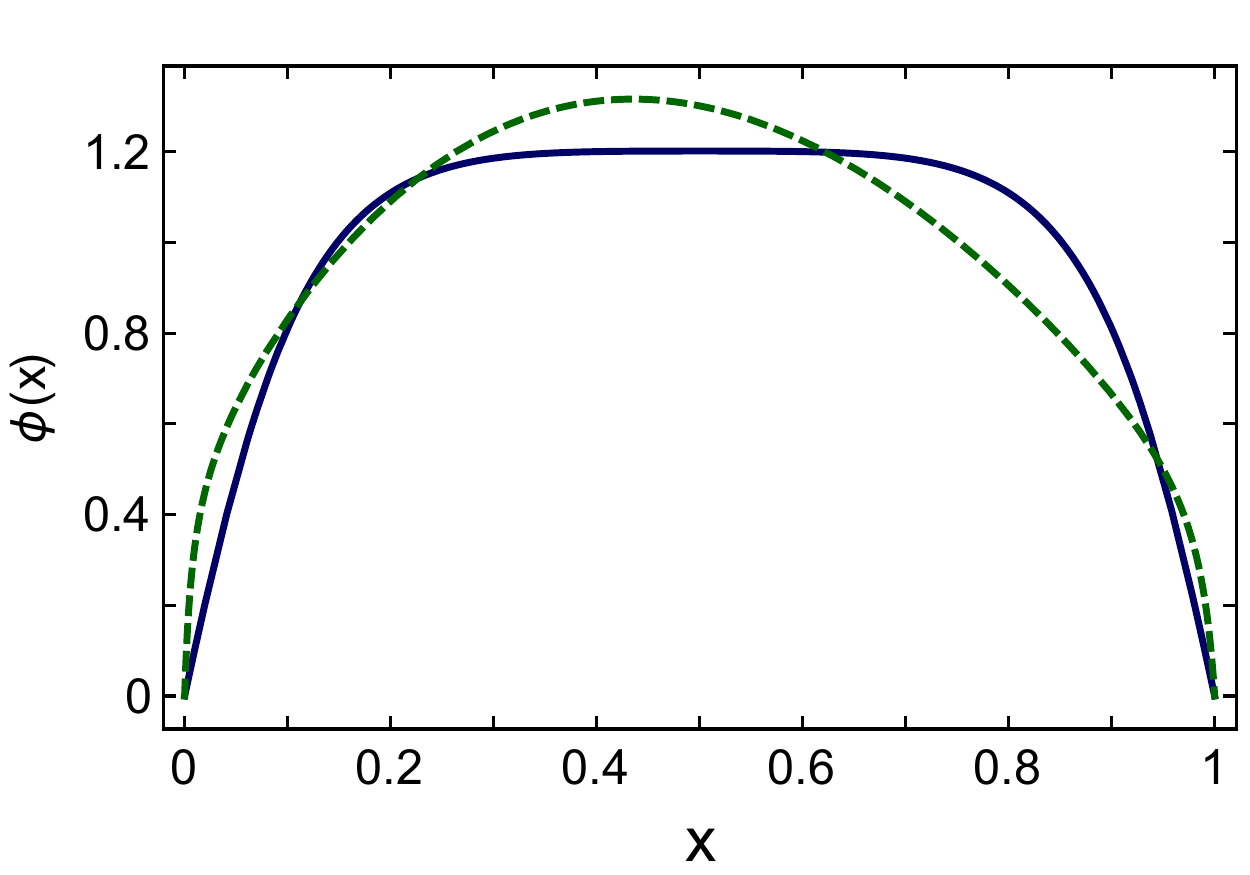}
\end{center}
\caption{\label{FigPDAs}
Leading-twist two-dressed-parton distribution amplitudes for the pion (solid, blue) and kaon (dashed, green) computed from the LFWFs in Fig.\,\ref{FigLFWF} using Eq.\,\eqref{EqVarphi}.}
\end{figure}

Having obtained the leading-twist LFWFs, one may compute the two-dressed-parton distribution amplitudes using Eq.\,\eqref{EqVarphi}, with the results depicted in Fig.\,\ref{FigPDAs}.  Consistent with the LFWFs, the PDAs are broad, concave functions.  Notably, although our models for the $\pi$ and $K$ are simple, they yield PDAs that agree qualitatively and semiquantitatively with results computed using more sophisticated approaches \cite{Chang:2013pq, Segovia:2013eca, Shi:2015esa, Horn:2016rip, Gao:2016jka, Gao:2017mmp}.  The peak in the kaon PDA lies at $x=0.44$.

\begin{figure}[t]
\begin{center}
\includegraphics[clip, width=0.43\textwidth]{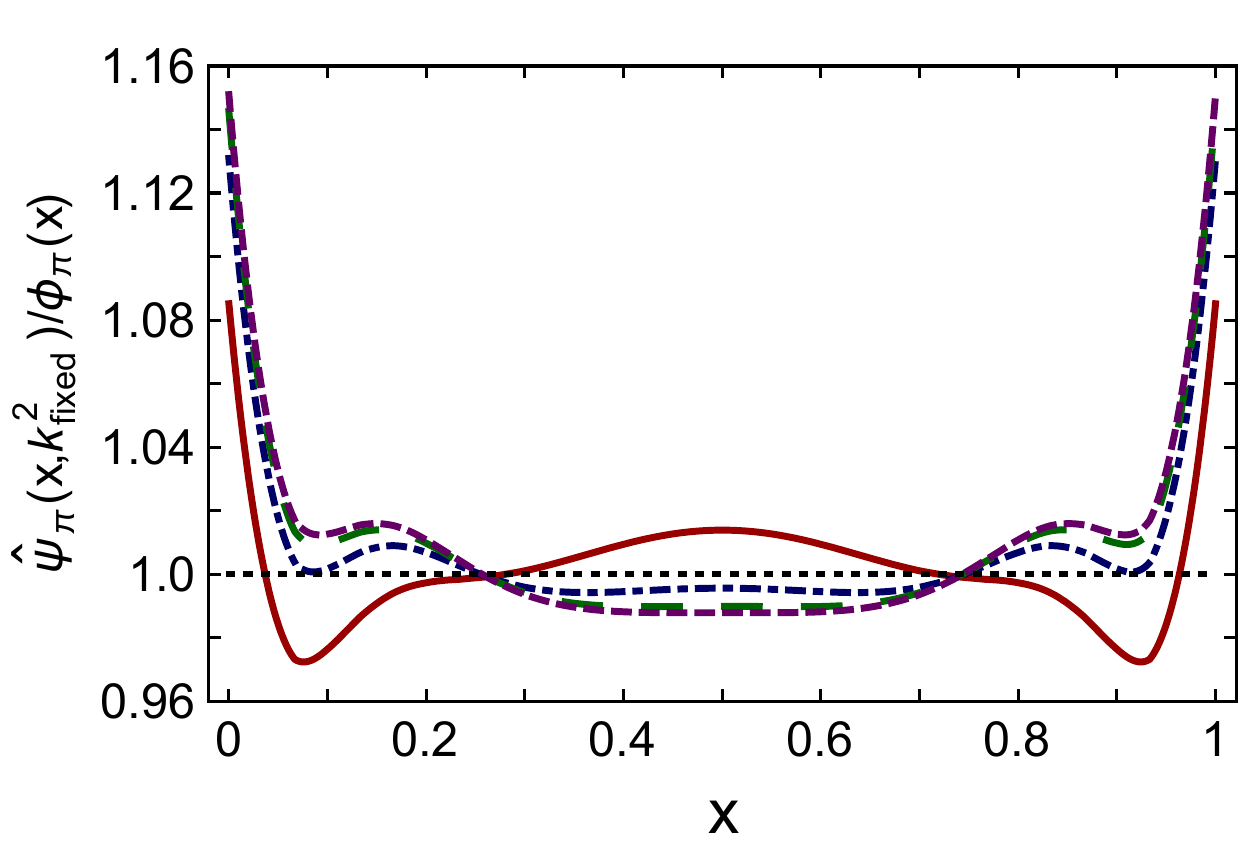}\hspace*{2ex}

\includegraphics[clip, width=0.42\textwidth]{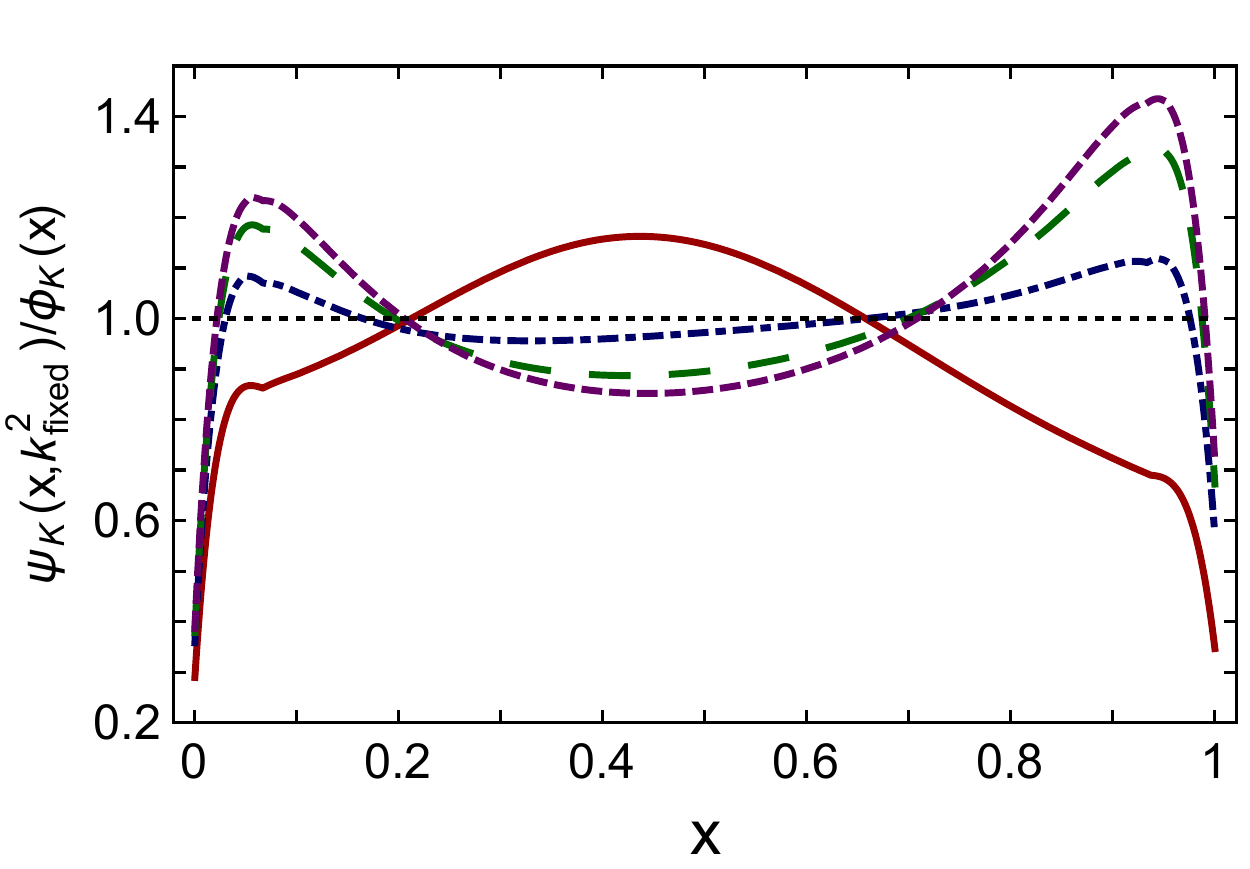}
\end{center}
\caption{\label{FigProduct}
\emph{Upper panel} -- $x$-dependence of the ratio in Eq.\,\eqref{EqRatio}, computed for the $\pi$-meson with $k_\perp^2/{\rm GeV}^2 =  0$ (solid, red), $0.2$ (dot-dashed, blue) $0.8$ (long-dashed, green), $3.2$ (dashed, purple).  %
\emph{Lower panel} -- same for kaon.
For a factorising wave function, this ratio would be unity: ${\mathpzc R}^{\psi\varphi}_G\equiv 1$, which is the dotted (black) line in both panels.}
\end{figure}

Notwithstanding the fact that a light-front Hamiltonian cannot in principle produce an eigenfunction of the form $\psi(x,k_\perp^2) \sim \psi_1(x) \psi_2(k_\perp^2)$ for any system because, \emph{e.g}.\ it violates momentum conservation, such a product \emph{Ansatz} is often used to produce numerical estimates of various quantities.  Given a hadron $G$, it is typically introduced in the form:
\begin{equation}
\label{EqFAnsatz}
\psi_G(x,k_\perp^2) \stackrel{\rm factorised\;\emph{Ansatz}}{=} \varphi_G(x) \, \psi_F(k_\perp^2)\,,
\end{equation}
with $\psi_F$, the $k_\perp^2$ profile function, often chosen to provide exponential decay.  In order to judge the accuracy of estimates obtained therewith, Fig.\,\ref{FigProduct} depicts the ratio:
\begin{equation}
\label{EqRatio}
{\mathpzc R}^{\psi\varphi}_G = \frac{\hat \psi_G^{\uparrow\downarrow}(x,k_\perp^2)}{\varphi_G(x)}\,,
\end{equation}
evaluated at a number of $k_\perp^2$-values and normalised at each such that $\int dx \, \hat \psi_G^{\uparrow\downarrow}(x,k_\perp^2) =1$.  If a product \emph{Ansatz} were a good approximation, then ${\mathpzc R}^{\psi\varphi}_G\equiv 1$.

The upper panel of Fig.\,\ref{FigProduct} depicts the ratio in Eq.\,\eqref{EqRatio} obtained for the pion, $G=\pi$.  It is not unity; and the discrepancy grows with increasing momentum until $k_\perp^2\approx 1\,$GeV$^2$, whereafter the ratio has a fairly static profile.  On the other hand, the departure from unity is not great: the ${\mathpzc L}_1$-deviation saturates at approximately $2$\%.  One might therefore argue that Eq.\,\eqref{EqFAnsatz}, with appropriate \emph{power-law} behaviour for $\psi_F(k_\perp^2)$, could be quantitatively useful for integrated properties of the pion and serve as a fair guide to the pointwise behaviour of $\psi_\pi(x,k_\perp^2)$.
One should nevertheless bear in mind that any product \emph{Ansatz} will be poorest on the domain of greatest correlation between the independent variables; and owing to momentum conservation, that domain is the neighbourhood of the endpoints, $x=0,1$, as evident in Fig.\,\ref{FigProduct}.

The situation is somewhat different for the kaon.  Depicted in the lower panel of Fig.\,\ref{FigProduct}, the ratio departs from unity by as much as 70\%.  The ${\mathpzc L}_1$-deviation is 15\% at $k_\perp^2=0$, initially drops with increasing $k_\perp^2$, but increases on $k_\perp^2 \gtrsim 0.1\,$GeV$^2$ to reach a limiting value of $\approx 20$\%.  In such circumstances, with a well-chosen power-law form for $\psi_F$,  Eq.\,\eqref{EqFAnsatz} might provide a fair indication of integrated kaon properties, but it can only at best be a sketchy guide to pointwise features of $\psi_K(x,k_\perp^2)$.


\begin{figure}[t]
\begin{center}
\includegraphics[clip, width=0.42\textwidth]{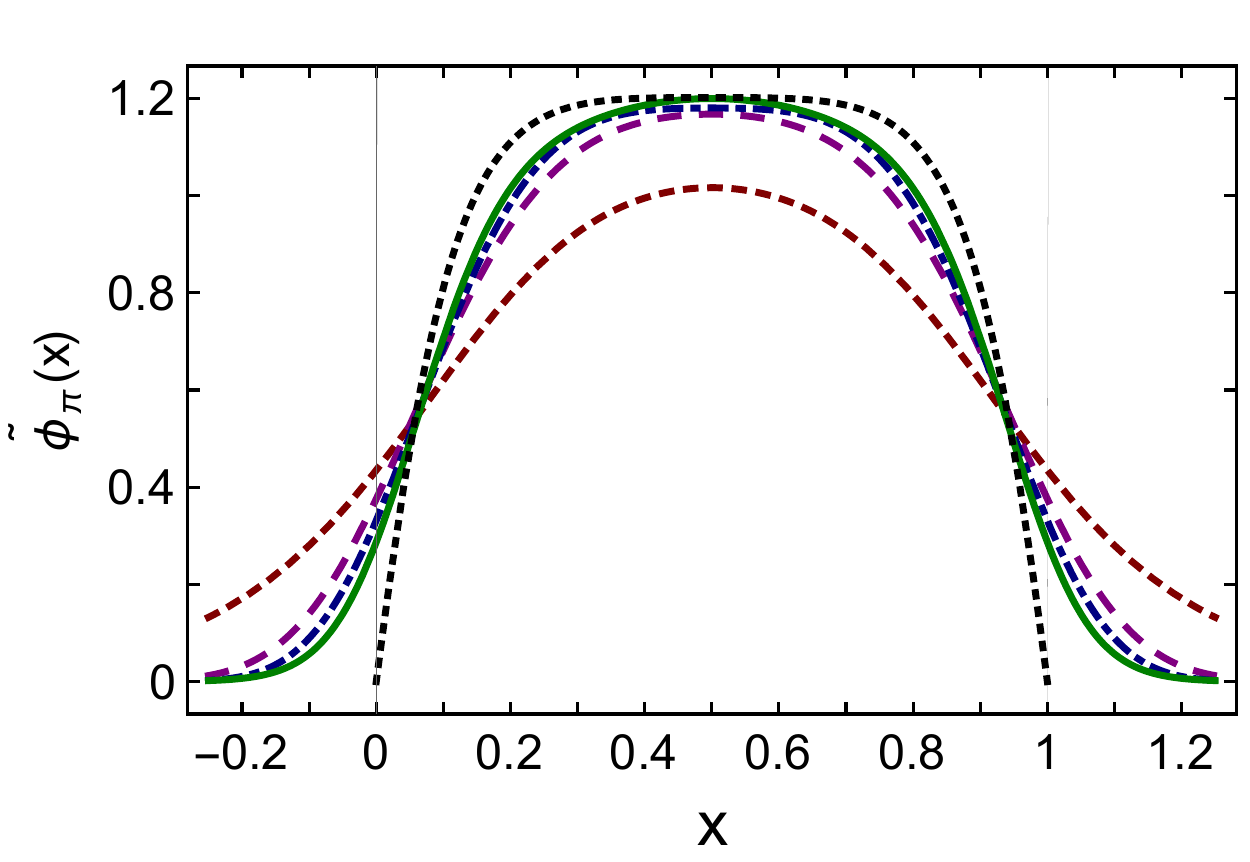}
\includegraphics[clip, width=0.42\textwidth]{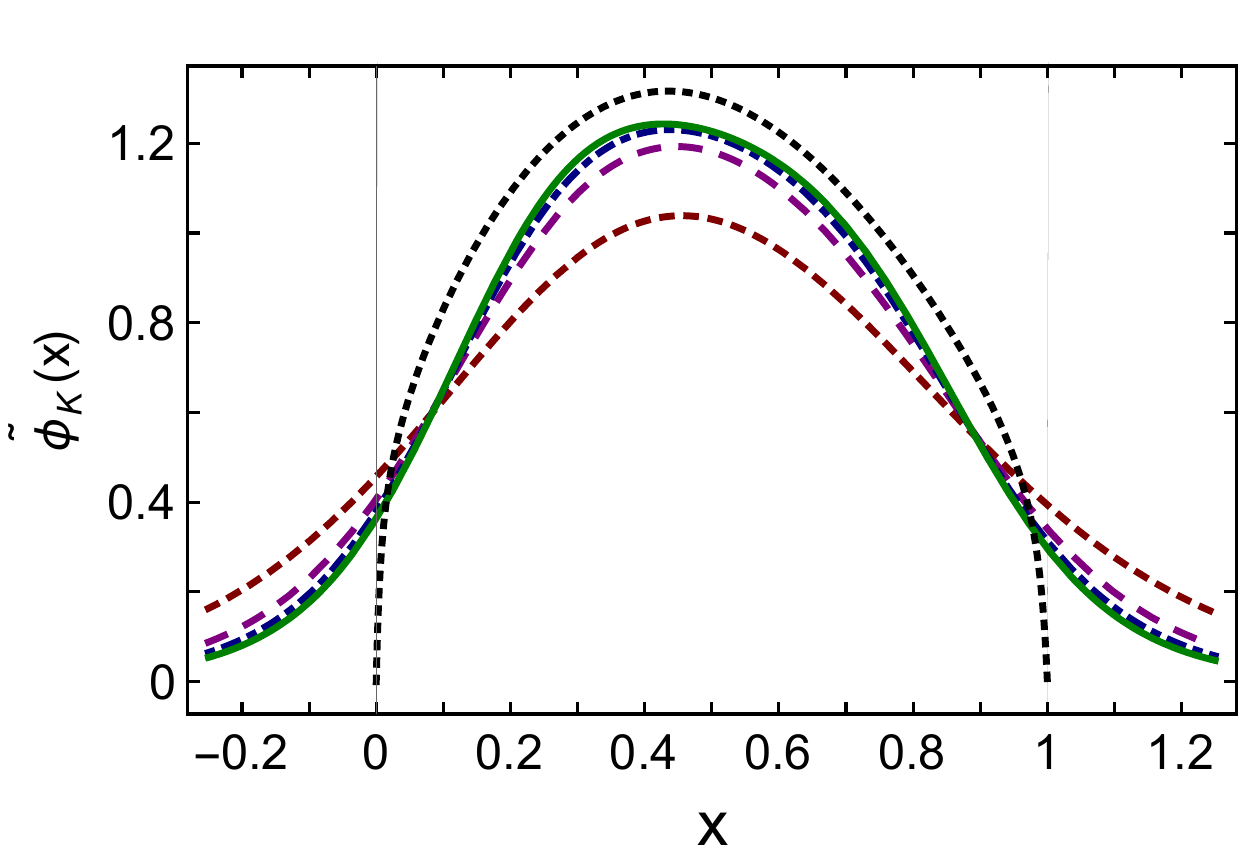}
\end{center}
\caption{\label{FigqPDApi}
\emph{Upper panel} -- $x$-dependence of the pion's quasi-PDA, computed with $P_z/{\rm GeV}=1$ (short-dashed, red), $1.75$ (dashed, purple), $2.4$ (dot-dashed, blue), $3.0$ (solid, green).
\emph{Lower panel} -- same for kaon.
The dotted (black) curve in both panels is the appropriate PDA from Fig.\,\ref{FigPDAs}; and the thin vertical lines at $x=0,1$ highlight the boundaries of support for a physical PDA.
}
\end{figure}

\subsubsection{Quasi Parton Distribution Amplitudes}
Eq.\,\eqref{EqqPDA} can now be used to compute pion and kaon qPDAs, with the results depicted in Fig.\,\ref{FigqPDApi}.
Focusing first on the pion (upper panel), it is evident that the result obtained with $P_z=1.0\,$GeV does not closely resemble $\varphi_\pi(x)$: the ${\mathpzc L}_1$-difference between the two curves is 42\% and the $(2x-1)^2$-moment obtained by integrating $\tilde \varphi_\pi(x)$ on $x\in [0,1]$ is just 33\% of the objective value.\footnote{The objective value for this moment is $0.25$, Eq.\,\eqref{EqMoments}.  On physical grounds \cite{Segovia:2013eca}, the pion's $(2x-1)^2$-moment should lie between the conformal limit value, $1/5$, and the result obtained using $\varphi_\pi(x)=\,$constant, \emph{viz}.\ $1/3$.  Using $\tilde\varphi_\pi(x;P_z=1\,{\rm GeV})$, the moment defined here takes the value $0.22$: $(0.22-1/5)/(0.25-1/5) = 0.33$.}

The step to $P_z=1.75\,$GeV brings material improvement, so that the $\tilde\varphi_\pi(x)$ provides a qualitatively sound approximation to $\varphi_\pi(x)$: the ${\mathpzc L}_1$-difference between the two curves is 18\%, the $(2x-1)^2$-moment is $78$\% of the objective value, and one can reasonably conclude that the target PDA is a broad, concave function.

Further increments in $P_z$, however, do not bring much improvement.  For example, with $P_z=3.0\,$GeV, the ${\mathpzc L}_1$-difference between the qPDA and the PDA is 10\% and the $(2x-1)^2$-moment is $85$\% of the objective value.  This outcome is a reflection of the fact that once the perturbative domain is entered, evolution in QCD is logarithmic.

It is noteworthy, too, that the pointwise forms of $\tilde\varphi_\pi(x)$ leak significantly from the domain $0<x<1$.  This prevents a determination of the target PDAs endpoint behaviour even with $P_z=3\,$GeV.  That behaviour is crucial because it fixes the magnitude of the leading-order, leading-twist perturbative QCD results for numerous observables \cite{Lepage:1980fj} and hence sets the benchmark against which existing and foreseen experiments aimed at testing solid QCD predictions must be compared \cite{Brodsky:2011yv, Chang:2013nia, Raya:2015gva, Raya:2016yuj, Gao:2017mmp}.
Notably, in order to reach $P_z \approx 3\,$GeV in a lQCD simulation, one would need a lattice with roughly $48$ spatial sites and a spacing of $0.06\,$fm.

Turning attention now to the kaon qPDAs, there are similarities with the pion case.  Using $P_z=1.75\,$GeV, $\tilde\varphi_K(x)$ provides some reliable qualitative information about $\varphi_K(x)$: the ${\mathpzc L}_1$-difference between the two curves is 28\%, the qPDA peaks at $x=0.45$, and its $(2x-1)^1$ moment is 75\% of the objective value.  On the other hand, the  $(2x-1)^2$-moment is just $29$\% of the goal.  Once again, incrementing $P_z$ does not greatly improve the situation.  Using $P_z=3.0\,$GeV, the ${\mathpzc L}_1$-difference between $\tilde\varphi_K(x)$ and $\varphi_K(x)$ is 20\%, the qPDA peaks at $x=0.43$, the $(2x-1)^1$ moment is 82\% of the objective value, but the  $(2x-1)^2$-moment is only $35$\% of the goal.  In this case, reducing the ${\mathpzc L}_1$-difference between $\tilde\varphi_K(x)$ and $\varphi_K(x)$ to 10\% would require $P_z \approx 20\,$GeV.

\begin{figure}[t]
\begin{center}
\includegraphics[clip, width=0.47\textwidth]{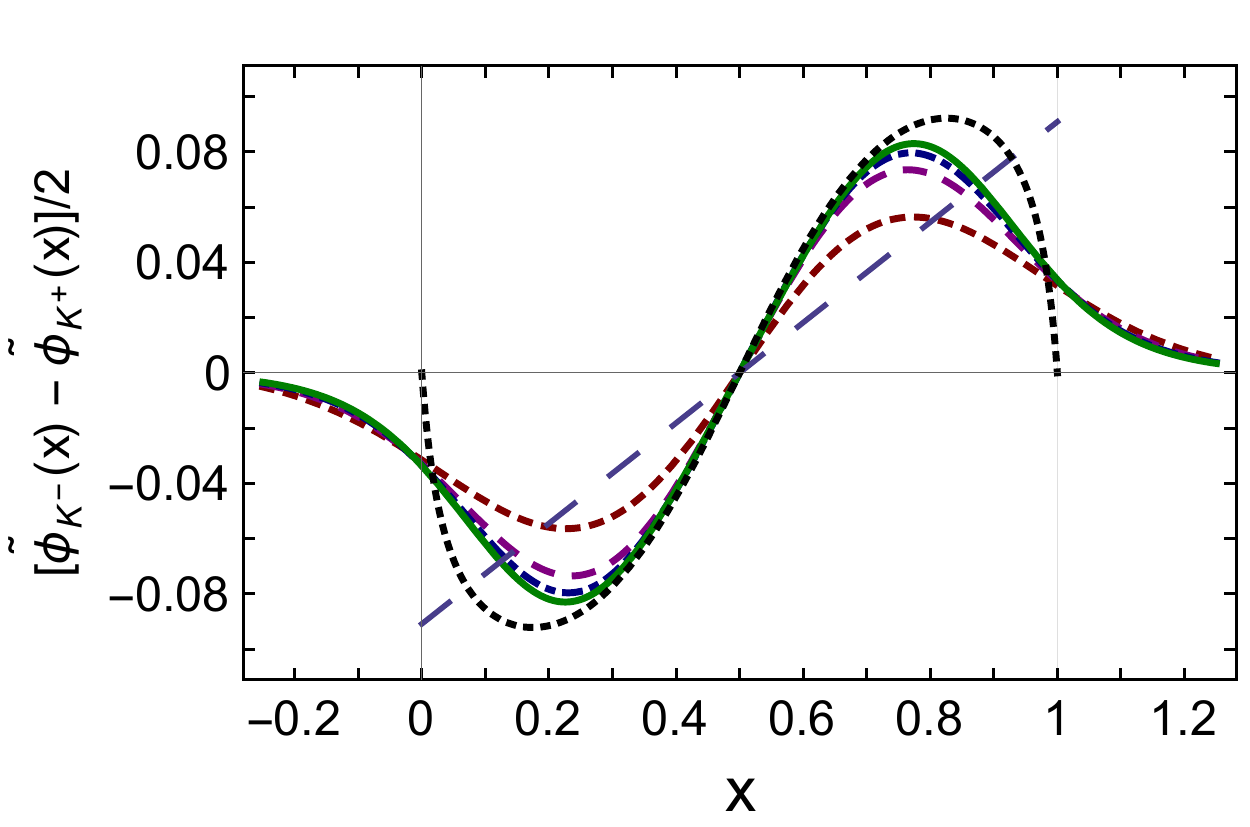}
\end{center}
\caption{\label{FigSkew}
$\tilde\varphi_{K^-}(x)-\tilde\varphi_{K^+}(x)$, computed with $P_z/{\rm GeV}=1$ (short-dashed, red), $1.75$ (dashed, purple), $2.4$ (dot-dashed, blue), $3.0$ (solid, green).
The dotted (black) curve is the result obtained with the objective $K^\pm$ PDAs; the long-dashed (slate-blue) curve is the function $\delta {\mathpzc M}(x)$ in Eqs.\,\eqref{MuMs}; and the thin vertical lines at $x=0,1$ highlight the boundaries of support for a physical PDA.
}
\end{figure}

In closing this subsection we return to $SU(3)$-flavour-symmetry violation in the kaon qPDAs, plotting $[\tilde\varphi_{K^-}(x)-\tilde\varphi_{K^+}(x)]/2$ in Fig.\,\ref{FigSkew}.  Evidently, using modest values of $P_z$, the qPDAs provide a fair pointwise description of the true difference on $x\in[0.3,0.7]$.  Again, however, the behaviour on large domains near the endpoints is poorly represented.  The figure also displays
\begin{subequations}
\label{MuMs}
\begin{align}
\label{MuMsE}
\delta {\mathpzc M}(x) & =\frac{ (M_u - M_s) (1-x) + x (M_s - M_u)}{M_u+M_s}\,, \\
& \approx \frac{ (f_\pi - f_K) (1-x) + x (f_K - f_\pi)}{f_\pi+f_K}\,.
\end{align}
\end{subequations}
The comparison of $\delta {\mathpzc M}(x)$ with $[\tilde\varphi_{K^-}(x)-\tilde\varphi_{K^+}(x)]$ highlights that the scale of flavour-symmetry breaking in the kaon distribution amplitudes measures differences between the emergent masses of $s$- and $u$-quarks in the Standard Model.  The analogue of Eq.\,\eqref{MuMsE} produced using Higgs-generated current-masses is an order-of-magnitude too large at the qPDAs' extrema.

\section{Quasi Parton Distribution Functions}
\label{SecqPDFSs}
\subsection{Algebraic Analysis: qPDFs}
In describing valence-dressed-quark parton distribution functions at an hadronic scale, $\zeta_H$, the  impulse-approximation (handbag diagram) is inadequate because it omits contributions from the gluons which bind valence-quarks into a hadron.  A remedy for this flaw is described and used to compute pion and kaon valence-quark distribution functions in Refs.\,\cite{Chang:2014lva, Chen:2016sno}.  Using the kaon as an illustration:
\begin{subequations}
\label{GPDFs}
\begin{align}
\nonumber
u_V^K(x) & = {\rm tr}_{\rm CD}\int_{dk} \delta_n^x(P_K) \\
& \quad \times [n\cdot \partial_k H_u(k;P_K)] H_s(k;P_K)\,,\\
s_V^K(x) & = u_V^K(1-x)\,,
\label{stouK}
\end{align}
\end{subequations}
where $n\cdot \partial_k = n_\mu (\partial / \partial k_\mu)$,
\begin{subequations}
\begin{align}
H_u(k;P_K) & = \bar \Gamma_K(k_-^K;-P_K) S_u(k)\,,\\
H_s(k;P_K) & = \Gamma_K(k_-^K; P_K) S_s(k-P_K)\,,
\end{align}
\end{subequations}
with $\bar\Gamma(k_-^K,P_K) = C^\dagger \bar\Gamma(-k_-^K,P_K)^{\rm T} C$, where $C$ is the charge conjugation matrix and $(\cdot)^{\rm T}$ denotes a transposed matrix.  Expressions for analogous distributions in the $\pi$ are obtained by changing $s\to d$.

Canonical normalisation of the kaon's Bethe-Salpeter amplitudes ensures
\begin{equation}
\label{EqNormPDF}
\int_0^1 dx \, u_V^K(x) = 1 = \int_0^1 dx \, s_V^K(x)\,.
\end{equation}
Consequently, using Eq.\,\eqref{stouK}, one finds immediately:
\begin{equation}
1 = \int_0^1 dx \, x [ u_V^K(x) + s_V^K(x)]\,.
\end{equation}

In obtaining these results, one must use mathematical features of the matrix trace, properties of propagators and Bethe-Salpeter amplitudes under charge conjugation, and the following identity: for $n^2=0$,
\begin{equation}
\label{0Identity}
0 = {\rm tr}_{\rm CD}\int_{dk} \delta_n^x(P_K) n\cdot \partial_k [H_u(k;P_K) \, H_s(k;P_K)]\,.
\end{equation}

Arriving at a quasi-PDF extension of Eqs.\,\eqref{GPDFs} is almost as straightforward as making the transition from PDAs to qPDAs, described in Sec.\,\eqref{SecqPDAs}: one has
{\allowdisplaybreaks
\begin{subequations}
\label{EqqPDFs}
\begin{align}
\nonumber
\tilde u_V^K(x) & = {\rm tr}_{\rm CD}\int_{dk} \delta_{\tilde n}^x(P_K) \\
& \times [\tilde n\cdot \partial_k H_u(k;P_K)] H_s(k;P_K) - {\mathpzc S}(x) \,,\\
\label{tildesVuV}
\tilde s_V^K(x) & = \tilde u_V^K(1-x)\,,\\
%
%
\nonumber
{\mathpzc S}(x) & = \tfrac{1}{2} {\rm tr}_{\rm CD}\int_{dk} \delta_{\tilde n}^x(P_K)  \\
& \quad \times \tilde n\cdot \partial_k [H_u(k;P_K) H_s(k;P_K)]\,.
\end{align}
\end{subequations}}
Analogous to the procedure in Sec.\,\eqref{SecqPDAs}, the primary step is simply $n\to \tilde n$ in the PDF formulae.  However, the correction term, ${\mathpzc S}(x)$, is also needed.  Its presence is suggested by the role of Eq.\,\eqref{0Identity} in ensuring momentum conservation; and it guarantees, \emph{inter alia}, Eq.\,\eqref{tildesVuV}.  Once again, analogous distributions in the $\pi$ are obtained by replacing $s\to d$.

Using Eqs.\,\eqref{EqqPDFs}, one may readily establish
\begin{subequations}
\begin{align}
\int_{-\infty}^\infty & d\tilde x\,  \tilde u_V^K(\tilde x) = \int_0^1dx\,u_V^K(x) = 1\,,\\
\int_{-\infty}^\infty & d\tilde x\,  \tilde{ s}_V^K(\tilde x)  = \int_0^1dx\, s_V^K(x) = 1\,, \\
\nonumber
\int_{-\infty}^\infty & d\tilde x \, \tilde x [ \tilde u_V^K(\tilde x) + \tilde  s_V^K(\tilde x)] \\
& \quad = \int_0^1 dx \, x [ u_V^K(x) + s_V^K(x)] = 1\,.
\end{align}
\end{subequations}
Evidently, Eqs.\,\eqref{EqqPDFs} define purely valence quark quasidistributions.

\subsection{Numerical Illustrations: qPDFs}
\begin{figure}[t]
\begin{center}
\includegraphics[clip, width=0.45\textwidth]{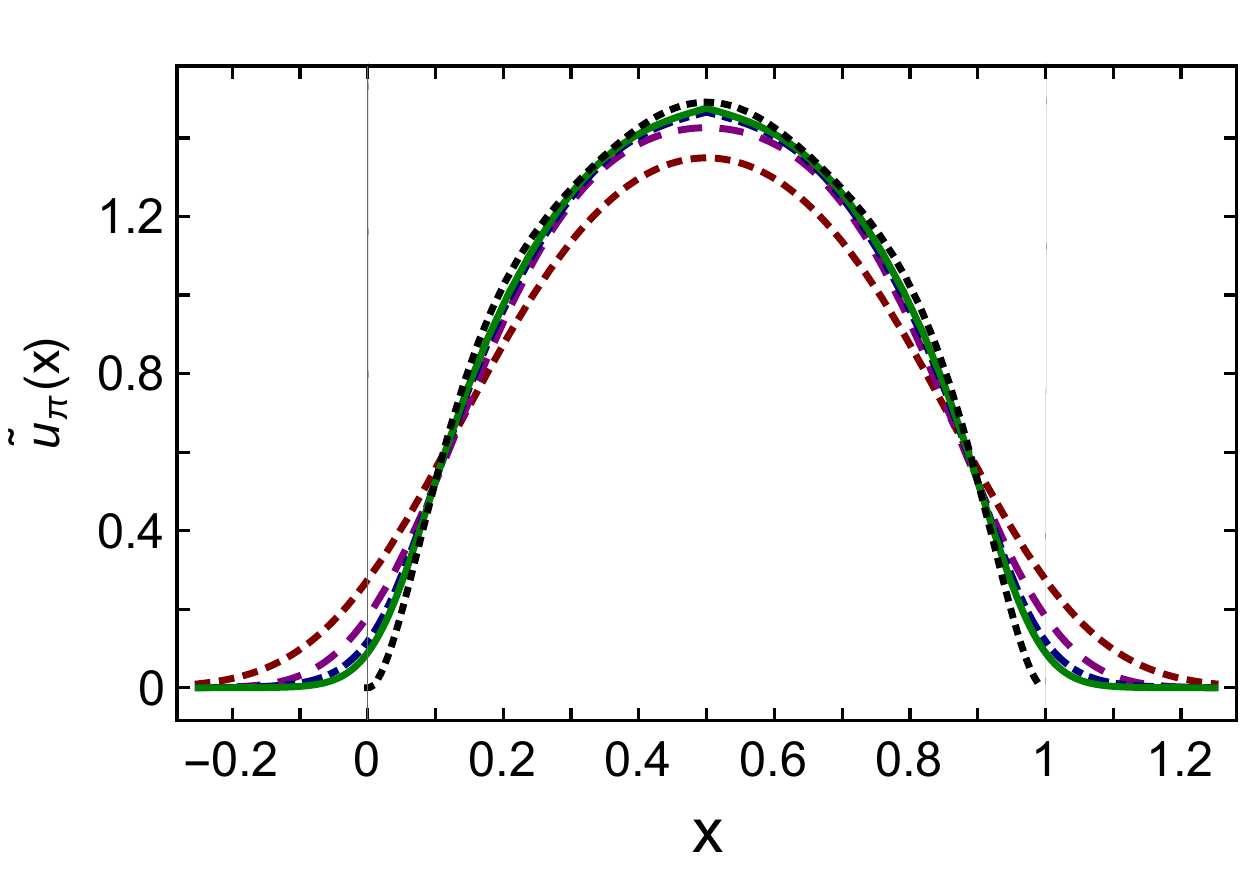}
\includegraphics[clip, width=0.45\textwidth]{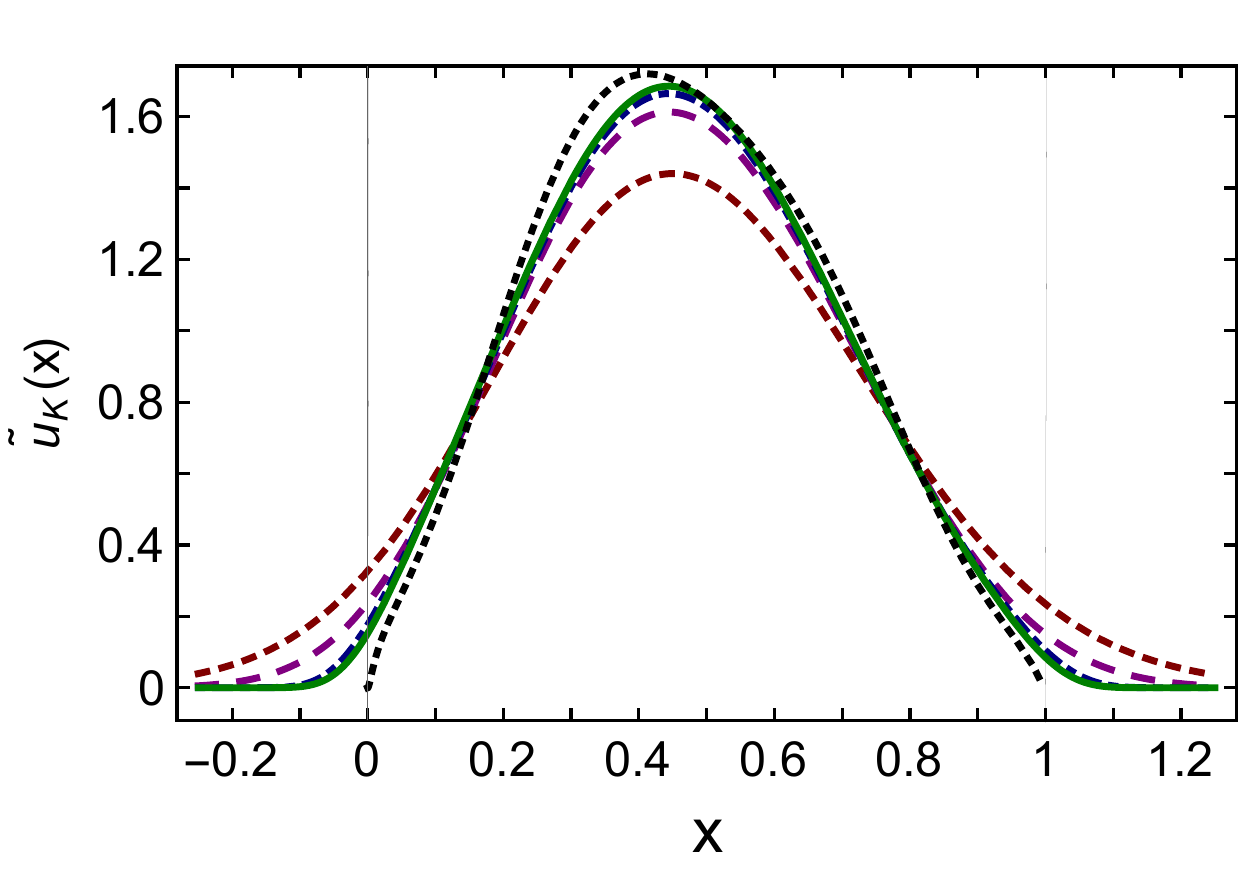}
\end{center}
\caption{\label{FigqPDFpi}
\emph{Upper panel} -- Pion's dres\-sed-valence $u$-quark quasi-PDF at the hadronic scale, computed with $P_z/{\rm GeV}=1$ (short-dashed, red), $1.75$ (dashed, purple), $2.4$ (dot-dashed, blue), $3.0$ (solid, green).
\emph{Lower panel} -- Same for kaon.
The dotted (black) curve in both panels is the associated objective PDF, computed using Eqs.\,\eqref{Algebraic}, \eqref{Eqrho}, \eqref{parameters}, \eqref{GPDFs}; and the thin vertical lines at $x=0,1$ highlight the boundaries of support for a physical valence-quark PDF.
}
\end{figure}

We now use Eqs.\,\eqref{Algebraic}, \eqref{Eqrho}, \eqref{parameters}, \eqref{GPDFs} to compute the pion and kaon qPDFs.  The calculation is straightforward, following the pattern in Sec.\,\ref{Algebraic}: one uses Feynman parametrisation to combine denominator products into a single quadratic form, Cauchy's theorem to evaluate the $k_4$ integral, direct evaluation for $\int d^2k_\perp$, and finally numerical integration over the Feynman parameters.  The results are depicted in Fig.\,\ref{FigqPDFpi}.  (The objective PDFs were obtained using the approach described in Ref.\,\cite{Chen:2016sno} and checked using the overlap representation \cite{Brodsky:1989pv}.)

A cursory comparison between Figs.\,\ref{FigqPDApi} and \ref{FigqPDFpi} reveals that a valence-quark qPDF is typically a better approximation to the objective result than a qPDA at any given value of $P_z$.  Looking closer at the pion (Fig.\,\ref{FigqPDFpi}, upper panel), the ${\mathpzc L}_1$-differences are
19\% ($P_z=1\,$GeV),
9\% ($P_z=1.75\,$GeV),
5\% ($P_z=2.4\,$GeV),
4\% ($P_z=3\,$GeV).
This series indicates that even with $P_z=1\,$GeV, the pion's valence-quark qPDF delivers a qualitatively sound approximation to the true result; and the step to $P_z=1.75\,$GeV brings noticeable improvement; but, as with the qPDAs, improvement is slow on $P_z > 1.75\,$GeV.

Similar, too, is the pointwise behaviour of the valence-quark qPDFs in the neighbourhood of the endpoints: as with the qPDAs, the qPDFs leak   significantly from the domain $0<x<1$.
This is important because one of the earliest predictions of the QCD parton model, augmented by features of perturbative QCD (pQCD), is that the valence-quark distribution function in a pseudoscalar meson behaves as follows \cite{Brodsky:1973kr, Brodsky:1974vy, Ezawa:1974wm, Farrar:1975yb, Berger:1979du, Ball:2016spl}:
\begin{equation}
\label{PDFQCD}
q_V^G(x;\zeta_H) \stackrel{{\rm large}\,x}{\sim} (1-x)^{2+\gamma},
\end{equation}
where $\gamma\gtrsim 0$ is an anomalous dimension.  Verification of Eq.\,\eqref{PDFQCD} is an important milestone on the path toward confirmation of QCD as the theory of strong interactions \cite{Holt:2010vj}.  In this connection we recall that Ref.\,\cite{Conway:1989fs} (the E615 experiment) reported a pion valence-quark PDF obtained via a leading-order pQCD analysis of their data, \emph{viz}.\ $u_V^\pi(x) \sim (1-x)$, seemingly a marked contradiction of Eq.\,\eqref{PDFQCD}.  Subsequent computations using continuum methods appropriate to QCD bound-states \cite{Hecht:2000xa} confirmed Eq.\,\eqref{PDFQCD} and eventually prompted reconsideration of the E615 analysis, with the result that at next-to-leading order and including soft-gluon resummation \cite{Wijesooriya:2005ir, Aicher:2010cb}, the E615 data can be viewed as being consistent with Eq.\,\eqref{PDFQCD}.  New data are essential in order to check this reappraisal of the E615 data and settle the controversy.  This goal is a focus of an approved tagged DIS experiment at the Thomas Jefferson National Accelerator Facility (JLab) \cite{Keppel:2015, McKenney:2015xis, R.A.Montgomery:2017hab}.  Such data could also be obtained with the common muon proton apparatus for structure and spectroscopy (COMPASS) detector at CERN \cite{Peng:2017ddf, Bourrely:2018yck} and at a future electron ion collider (EIC) \cite{Holt:2000cv, Horn:2018ghc}.

\begin{figure}[t]
\begin{center}
\includegraphics[clip, width=0.45\textwidth]{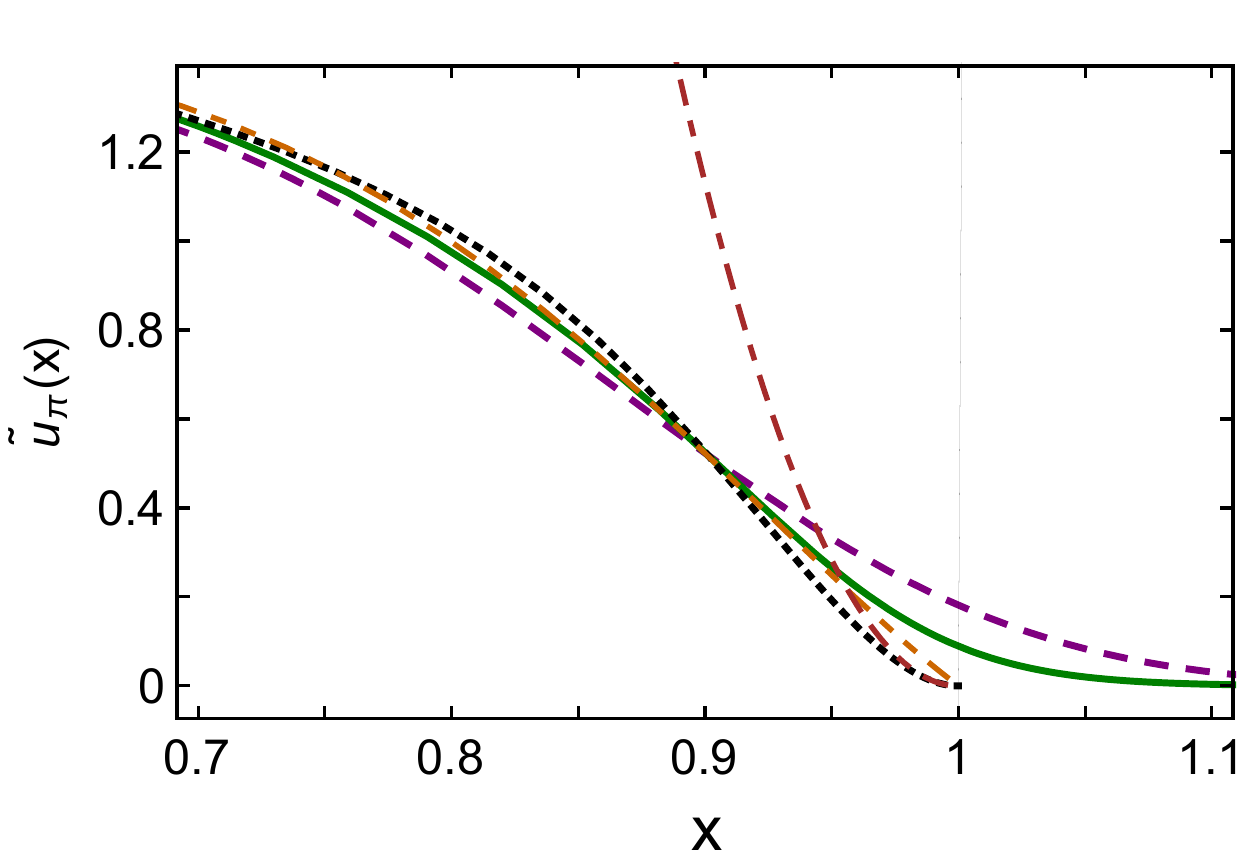}
\end{center}
\caption{\label{ValencePDF}
Pion qPDFs on $x>0.7$, \emph{i.e}.\ a valence-quark domain:
$P_z/{\rm GeV}=1.75$ (dashed, purple),
$3.0$ (solid, green).
Dotted (black) curve, objective PDF, $u_V^\pi(x)$, whose large-$x$ behaviour is given in Eq.\,\eqref{uvppilargex}; dot-dash-dashed (brown) curve, rhs of Eq.\,\eqref{uvppilargex}; and dot-dot-dashed (orange) curve, a PDF that is pointwise near-equivalent to $u_V^\pi(x)$, but which is $\propto (1-x)^1$ at large $x$.  (The thin vertical line at $x=1$ marks the upper bound on the domain of support for a physical valence-quark PDF.)
}
\end{figure}

These observations emphasise that quantitatively reliable lQCD results bearing upon Eq.\eqref{PDFQCD} would be very valuable. However, the challenge to delivering such outcomes using qPDFs is highlighted by Fig.\,\ref{ValencePDF}.  On a domain of valence-quark $x$, this figure compares the pion qPDFs in the upper panel of Fig.\,\ref{FigqPDFpi} with the objective-PDF:
\begin{equation}
\label{uvppilargex}
u_V^\pi(x)\stackrel{x>0.95}{\approx} 113\, (1 - x)^2,
\end{equation}
and another curve, whose $x\in[0,1]$ ${\mathpzc L}_1$-difference from the objective valence-quark PDF is just 2\%, but which is $\propto (1-x)^1$  at large-$x$.  Evidently, even the $P_z=3\,$GeV qPDF is unable to distinguish between these two distinctively different results.  (That $x>0.9$ is required before $(1-x)^2$ behaviour is visible in the pion's valence-quark distribution was remarked upon earlier \cite{Holt:2010vj}.)


We now redirect our attention to kaon valence-quark qPDFs.  As with kaon qPDAs, there are similarities with the pion.  For instance, ${\mathpzc L}_1$-differences are
25\% ($P_z=1\,$GeV),
12\% ($P_z=1.75\,$GeV),
8\% ($P_z=2.4\,$GeV),
6\% ($P_z=3\,$GeV),
indicating, again, that even with $P_z=1\,$GeV, the kaon's qPDF delivers a qualitatively sound approximation to $u_V^K(x)$; the step to $P_z=1.75\,$GeV brings noticeable improvement, but changes are slow thereafter.
The remarks made in connection with the pion qPDFs' large-$x$ behaviour hold with equal force for the kaon.

\begin{figure}[t]
\begin{center}
\includegraphics[clip, width=0.45\textwidth]{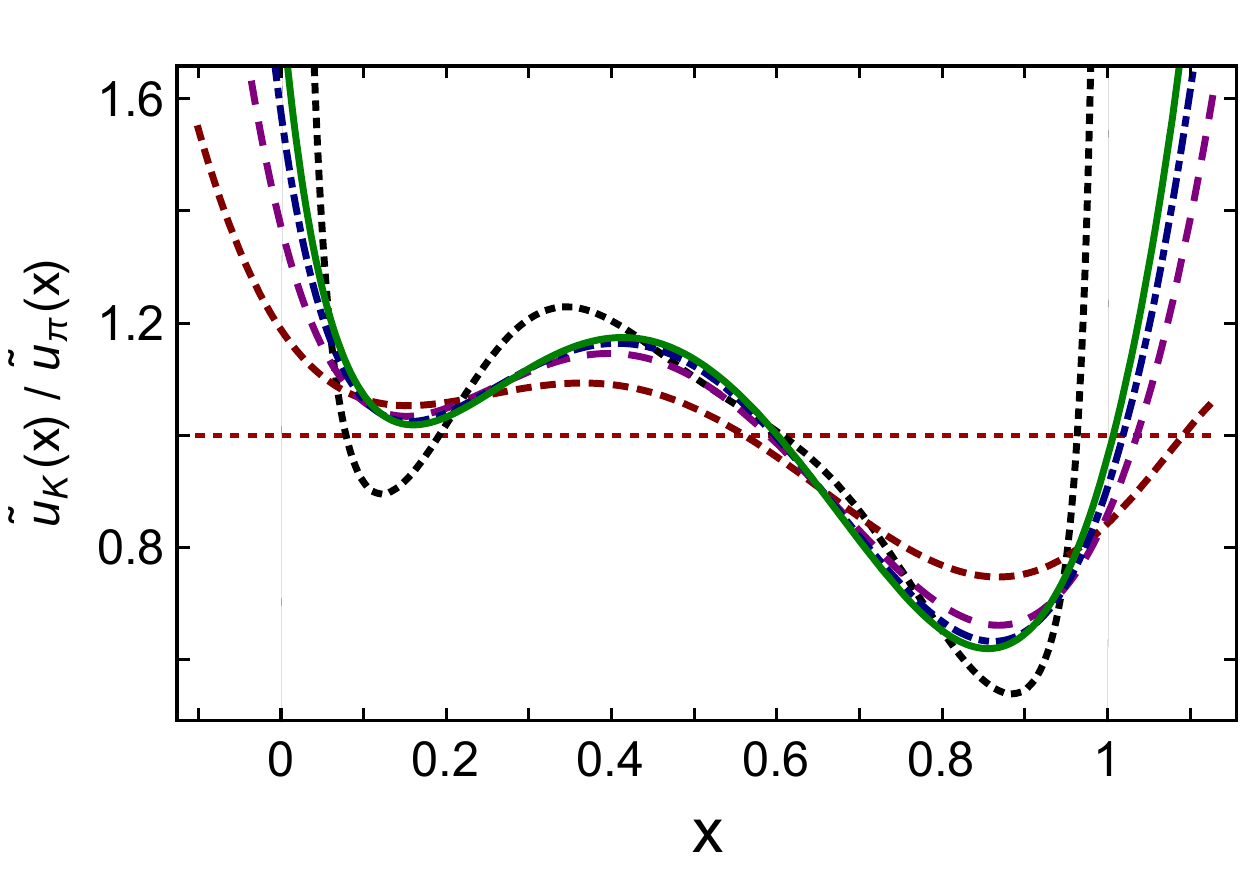}
\end{center}
\caption{\label{FigqPDFRatio}
$x$-dependence of the qPDF ratio $\tilde u_K/\tilde u_\pi$ at the hadronic scale, $\zeta_H$, computed with $P_z/{\rm GeV}=1$ (short-dashed, red), $1.75$ (dashed, purple), $2.4$ (dot-dashed, blue), $3.0$ (solid, green).
The dotted (black) curve is the associated objective ratio, $u_K/u_\pi$, obtained using the dotted (black) curves in Fig.\,\ref{FigqPDFpi}.
(The dotted (red) line is drawn at unity; and the thin vertical lines at $x=0,1$ highlight the boundaries of support for a physical valence-quark PDF.)
%
}
\end{figure}

It has been argued that the ratio $u_V^K(x)/u_V^\pi(x)$ serves as a sensitive probe of the difference between the gluon distributions in the pion and kaon \cite{Chen:2016sno}, and that this difference can reveal much about the emergence of mass in the Standard Model \cite{Roberts:2016vyn}.  Experimental data on the ratio is available \cite{Badier:1980jq}, but one measurement is insufficient for complete confidence.  Newer data would be welcome, in which connection tagged DIS at JLab might also be useful \cite{Horn:2016rip, C12-15-006A}, as could the COMPASS detector at the CERN  \cite{Peng:2017ddf, Bourrely:2018yck} and a future EIC \cite{Holt:2000cv, Horn:2018ghc}.  With these things in mind, in Fig.\,\ref{FigqPDFRatio} we depict the ratio $\tilde u_V^K(x)/\tilde u_V^\pi(x)$.
Evidently, for $P_z\geq 1.75\,$GeV, much as was the case with the qPDA asymmetry depicted in Fig.\,\ref{FigSkew}, the ratio of qPDFs is quantitatively a good approximation to the objective ratio on a material domain, \emph{viz}.\  $0.3\lesssim x \lesssim 0.8$.  This domain almost covers that upon which empirical data is available.  We therefore anticipate that contemporary lQCD simulations could provide a sound prediction for this ratio before next generation experiments are completed.

\section{Summary and Perspective}
\label{secEpilogue}
Employing a continuum approach to bound-states in quantum field theory and practical algebraic \emph{Ans\"atze} for the Poincar\'e-covariant Bethe-Salpeter wave functions of the pion and kaon, we computed the leading-twist two-dressed-parton light-front wave functions (LFWFs), $\psi(x,k_\perp^2)$; parton distribution amplitudes (PDAs), $\varphi(x)$; quasi-PDAs (qPDAs), $\tilde\varphi(x)$; valence parton distribution functions (PDFs), $u_V(x)$; and quasi-PDFs (qPDFs), $\tilde u_V(x)$, for these systems.

The LFWFs are broad, concave functions, with power-law $k_\perp^2$-decay.  Whilst the pion's LFWF is symmetric about $x=1/2$, $\psi_K(x,k_\perp^2)$ peaks at $(x=0.44,k_\perp^2=0)$, expressing $SU(3)$-flavour-symmetry violation with a magnitude determined by differences between dynamical (not explicit) mass generation in the $s$- and $u$-quark sectors.  Looking closely at the LFWFs, we found that a carefully constructed product \emph{Ansatz}, \emph{viz}.\ $\psi(x,k_\perp^2) \sim \psi_1(x) \psi_2(k_\perp^2)$, although flawed in principle, can provide fair estimates of integrated $\pi$, $K$ properties.

The LFWFs provide direct access to $\pi$ and $K$ PDAs and qPDAs; and for each system the qPDF provides a semiquantitatively reliable representation of the associated PDA when computed using a longitudinal momentum $P_z=1.75\,$GeV.  However, improvements thereafter are slow; and, notably, even with $P_z=3\,$GeV, the qPDA cannot provide information about the true PDAs endpoint behaviour.

Regarding pion and kaon valence-quark PDFs and qPDFs, we found that at any given $P_z$, a qPDF delivers a better representation of the associated PDF than does a qPDA of the objective PDA.  In fact, even with $P_z = 1\,$GeV the qPDF provides a qualitatively clear picture of the PDF.  However, as with qPDAs, differences between qPDFs and PDFs diminish slowly on $P_z>1.75\,$GeV; and, similarly, even with $P_z=3\,$GeV, qPDFs cannot be used to determine the objective PDF's large-$x$ behaviour.  On the other hand, the ratio $\tilde u_V^K(x)/\tilde u_V^\pi(x)$ does provide a good approximation to $u_V^K(x)/u_V^\pi(x)$ on $0.3\lesssim x \lesssim 0.8$, in consequence of which we expect that contemporary simulations of lattice-regularised QCD can deliver a reasonable prediction for this ratio before next generation experiments are completed.

It is natural to extend this analysis to the neutron and proton, for which analogous algebraic \emph{Ans\"atze} for the bound-state Faddeev wave functions exist or can readily be developed \cite{Mezrag:2017znp}.  We anticipate that the outcome will be qualitatively similar: in particular, that even using modest values of $P_z$, a material valence-quark $x$-domain will exist upon which ratios of qPDFs may provide sound representations of the PDF ratios measured empirically \cite{Hawker:1998ty, Baillie:2011za, Tkachenko:2014byy, Reimer:2016dcd}.


%
\acknowledgments
We are grateful for constructive suggestions from
J.~Chen, T.~Horn, L.~Liu, C.~Mezrag, J.~Rodr{\'i}guez-Quintero and S.~Riordan.
Work supported by:
China Postdoctoral Science Foundation (under Grant No.\ 2016M591809);
the Chinese Government's \emph{Thousand Talents Plan for Young Professionals};
the Chinese Ministry of Education, under the \emph{International Distinguished Professor} programme;
U.S.\ Department of Energy, Office of Science, Office of Nuclear Physics, under contract no.~DE-AC02-06CH11357;
and National Natural Science Foundation of China (under Grant Nos.\ 11475085, 11535005 and 11690030).


\end{document}